\documentclass[prl,twocolumn,floatfix,superscriptaddress]{revtex4}
\usepackage{graphicx}
\usepackage{amssymb}

\begin{document}
\title{Correlation effects on resonant tunneling in 
  one-dimensional quantum wires}
\author{V.\ Meden}
\affiliation{Inst.\ f.\ Theoret.\ Physik, Universit\"at G\"ottingen, 
Friedrich-Hund-Platz 1, D-37077 G\"ottingen, Germany}
\author{T.\ Enss}
\affiliation{Max-Planck-Institut f\"ur Festk\"orperforschung,
  Heisenbergstr.\ 1, D-70569 Stuttgart, Germany}
\author{S.\ Andergassen}
\affiliation{Max-Planck-Institut f\"ur Festk\"orperforschung,
  Heisenbergstr.\ 1, D-70569 Stuttgart, Germany}
\author{W.\ Metzner}
\affiliation{Max-Planck-Institut f\"ur Festk\"orperforschung,
  Heisenbergstr.\ 1, D-70569 Stuttgart, Germany}
\author{K.\ Sch\"onhammer}
\affiliation{Inst.\ f.\ Theoret.\ Physik, Universit\"at G\"ottingen, 
Friedrich-Hund-Platz 1, D-37077 G\"ottingen, Germany}

\begin{abstract}
We study resonant tunneling 
in a Luttinger liquid with a double 
barrier enclosing a dot region. Within a microscopic model calculation  
the conductance $G$ as a function of temperature $T$ is determined 
over several decades. 
We identify parameter regimes in which the peak value $G_{p}(T)$ 
shows distinctive power-law behavior. 
For intermediate dot 
parameters $G_p$ behaves in a non-universal way.
\end{abstract}
\maketitle     

A decade ago resonant tunneling through a double barrier 
embedded in a Luttinger liquid was intensively studied 
theoretically.\cite{KaneFisher,Furusaki1,Chamon} 
At temperature $T=0$, in the limit of an infinite system, and for symmetric 
barriers the conductance $G$ was predicted to show a series of infinitely sharp
resonances if the gate voltage $V_{ g}$ applied to the dot region
is varied.\cite{KaneFisher} 
For spinless fermions, on which we focus, the
peak conductance $G_{p}$ at resonance voltage $V_{g}^{r}$
is given by $e^2/h$ if the interacting wire
(denoted ``the wire'' in the following) is connected by ``perfect'' 
contacts to non-interacting semi-infinite leads as will be the
case here.
Increasing $T$ resonances are still present but acquire a non-vanishing 
width $w(T) \propto T^{1-K}$,\cite{KaneFisher} with  the interaction 
dependent Luttinger liquid 
parameter $K$ ($K=1$ for non-interacting particles). 
We here restrict ourselves to the case 
$1/2 \leq K \leq 1$. For $\Delta V_{ g} = |V_{ g} - V_{ g}^{r}| \neq 0$, 
$G(T)$ scales to zero as $T^{2 \alpha_{ B}}$,\cite{KaneFisher}
with $\alpha_B=1/K-1$. 
This also holds for an asymmetric double barrier regardless
if the conductance is evaluated at or off resonance.

Using second order perturbation theory in the barrier height the 
$T>0$ deviations of $G_p(T)$ from $e^2/h$ were determined to 
scale as $T^{2K}$.\cite{Furusaki1} 
At further increasing $T$ a regime of
uncorrelated sequential tunneling (UST) was predicted\cite{Furusaki2} 
based on a perturbative analysis in the inverse barrier height, 
where $G_{ p}(T) \propto T^{\alpha_{ B}-1}$ and $w(T) \propto T$.
This regime is bounded from above by the level 
spacing of the dot $\Delta_D=\pi v_{ F}/N_D$, where $v_{ F}$ 
denotes the Fermi velocity, $N_D$ the number of lattice sites 
in the dot, and the lattice constant was chosen to be 1. 
Within the same perturbative approach and for 
$\Delta_D < T < B$, where $B$
is the bandwidth, $G_p(T)$ increases as $T^{2 \alpha_B}$ 
for increasing $T$.\cite{Furusaki1,Furusaki2}  
In contradiction to the temperature 
dependence following from UST, in transport experiments on carbon nanotubes a 
suppression of  $G_p$ with decreasing $T$ was observed for 
$T \lessapprox \Delta_D$.\cite{Postma}  
Using another approximation scheme  
``correlated sequential tunneling'' (CST) was predicted to replace 
UST.\cite{Thorwart} It leads to $G_{ p} \propto T^{2
  \alpha_{ B}-1}$ which was argued to be consistent
with the experimental data. This consideration has stimulated a number of
theoretical works.\cite{Komnik,Nazarov,Polyakov,Huegle} 

Quantum Monte Carlo (QMC) results\cite{Huegle} for 
$G_p(T)$ were 
interpreted to be consistent with CST for $T \lessapprox \Delta_D$.  

In Refs.\ \cite{Nazarov,Polyakov} a  ``leading-log'' type of
resummation for the effective transmission at the chemical potential
was applied and led to important insights. 
In this approach 
no signature of CST was found, but it was criticized for only being 
valid in the limit of small interaction. That the ``leading-log''
method indeed does not capture all interesting interaction effects 
away from $K \to 1$ can be seen considering the single barrier case 
for which the conductance  for different temperatures and barrier 
heights can be collapsed onto a single 
curve by using a one-parameter scaling ansatz.\cite{KaneFisher} 
While the resulting
scaling function is known to depend on $K$,\cite{KaneFisher} the
``leading-log'' approach leads to the non-interacting $K=1$ scaling
function independently of the interaction strength 
chosen.\cite{Glazman,VM1}    

In Ref.\ \cite{Komnik} a $N_D=1$ double barrier model in which chiral
fermions carry the current was treated for $K=1/2$ by
refermionization. After fine tuning the Coulomb coupling across
the barriers $G_p(T)$ was predicted to show the 
non-interacting behavior, i.e.\ $e^2/h-G_p(T) \propto T^2$ for 
$T \to 0$ and $G_p(T) \propto T^{-1}$ in the UST regime. 
This finding is surprising and not supported by any of the 
results obtained for more general models using the above mentioned 
methods. The physics of this particular model thus seems to be 
non-generic.   

We study the problem applying the fermionic functional renormalization group
(fRG) method to the model of spinless fermions on a 
lattice with hopping $t=1$ and nearest neighbor interaction $U>0$ at 
half filling.\cite{VM1,VM2,SA} By using a method
which  can be applied for any barrier height
and shape  but is restricted to weak to intermediate interactions 
our study is complementary to the one   
using perturbation theory in either the barrier heights or the 
inverse barrier heights.\cite{Furusaki1,Furusaki2} 
Considering the single barrier case and comparing with exact 
asymptotic and numerical results we have shown previously that 
our method leads to meaningful results even for fairly large 
$U \lessapprox 2$.\cite{VM1,VM2,SA} In Ref.\ \cite{VM1} 
it was shown that for $K=1/2$ our data for $G$ 
collapse onto the exactly known scaling function of the local 
sine-Gordon model.\cite{KaneFisher} Without any further 
approximations the fRG can be applied to the double barrier 
problem and we thus expect that regimes with universal power-law 
scaling and the related exponents can reliably be determined.   
Accurate results can also be obtained for 
regions of dot parameters not investigated so far in 
which $G_p(T)$ shows non-universal behavior, i.e.\ a behavior 
which cannot be characterized
as a power-law scaling with an exponent expressible in terms of
$K$. Studying these turns out to be relevant  for 
the interpretation of the QMC data\cite{Huegle} and the experimental
results.\cite{Postma} 
In contrast to the other methods which can only be applied in 
certain temperature ranges the fRG provides reliable results for $G(T)$ 
on all energy scales.
For large barriers and large as well as small dots we identify a variety of 
temperature regions in which $G_{ p}(T)$ shows 
distinctive power-law behavior with different exponents 
expressible in terms of $K$. 
For intermediate barrier height, intermediate dot sizes, and  
$T  \lessapprox \Delta_D$, $G_{p}(T)$ decreases with decreasing $T$. 
The behavior of $G_p(T)$ in this temperature regime depends on 
$N_D$ and the height of the barriers and is thus non-universal. 
Whereas our results are consistent with 
the QMC data, we do not observe signatures of CST.

The model is given by the Hamiltonian 
\begin{eqnarray}
H  =  - \sum_{j=-\infty}^{\infty} 
\left( c_j^{\dag} c_{j+1}^{} +
  \mbox{h.c.} \right) + V_l n_{j_l-1} + V_r n_{j_r+1} \nonumber  \\ 
+ V_{ g} \sum_{j=j_l}^{j_r} n_j   + 
\sum_{j=1}^{N-1} U_j \left(n_j- \frac{1}{2} \right)  
\left(n_{j+1}- \frac{1}{2}\right) \; , 
\label{spinlessfermdef}
\end{eqnarray}
with $1 \ll  j_l < j_r \ll N$,
in standard second-quantized notation.
The lattice can be divided in three parts: (1) the wire with nearest
neighbor interaction $U_j$ across the bonds $(j,j+1)$ 
with $j \in [1,N-1]$; 
(2) the dot (embedded in the wire) on sites 
$j \in [j_l,j_r]$ 
($N_D = j_r-j_l+1$) which can be shifted in energy by the onsite 
energy $V_g$ and is separated by barriers of strength $V_{l/r}$ 
from the rest of the lattice; (3) the non-interacting leads on sites 
$j < 1$ and $j> N$. 
Besides site impurities we also consider  
hopping impurities as barriers. 
In this case the interaction [in the last term of Eq.\
(\ref{spinlessfermdef})] and hopping [in the first term of Eq.\
(\ref{spinlessfermdef})] across the bonds $j_l-1,j_l$ and 
$j_r,j_r+1$ are set to zero. Furthermore the second and third 
terms in the first line of Eq.\
(\ref{spinlessfermdef}) are replaced by 
$-t_l c_{j_l-1}^{\dag} c_{j_l}^{} - t_r c_{j_r}^{\dag} c_{j_r+1}^{} +
\mbox{h.c.}$ with the reduced hopping matrix elements $t_{l/r} < 1$. 
As in the other theoretical studies mentioned
we avoid effects of the contacts. 
We thus model ``perfect''  contacts by a smooth spatial variation 
of the interaction $U_j$ close to the sites $1$ and $N$. 
Starting at site $1$ in our calculations the interaction was turned on 
in an $\arctan$-shape over the first 100 sites and similarly turned 
off approaching site $N$.\cite{VM1} For $T=0$, vanishing barriers, and 
up to $N=10^4$ this gives $1- G/(e^2/h) < 10^{-6}$. The 
constant bulk value of the interaction is denoted by $U$. 

In linear response and the absence of
vertex corrections the conductance is given by\cite{Oguri}
\begin{eqnarray}
G(T) = - (e^2/h) \int_{-2}^{2}
 |t(\varepsilon,T)|^2  \; \partial_{\varepsilon} f(\varepsilon/T) 
\; d \varepsilon 
\, ,
\label{conductf}
\end{eqnarray} 
with $|t(\varepsilon,T)|^2 = (4- \varepsilon^2) 
|\tilde G_{1,N}(\varepsilon,T)|^2$ and $f$ being the
Fermi function. 
The one-particle Green function $\tilde G_{i,j}$ has to be
computed in the presence of interaction and leads. To this
end we extend a recently developed fRG scheme, which has been
shown to provide excellent results at $T=0$,\cite{VM1,VM2,SA} 
to finite temperatures. The starting point of this approach is an exact
hierarchy of differential flow equations for the self-energy 
$\Sigma^{\Lambda}$
and higher order vertex functions, with an energy cutoff $\Lambda$
as the flow parameter. The hierarchy is truncated by neglecting
$n$-particle vertices with $n>2$, and the 2-particle vertex
is parametrized approximately by a renormalized nearest
neighbor interaction $U^{\Lambda}$. The spatial dependence of the
renormalized impurity potential, described by 
$\Sigma^{\Lambda}$, is however fully kept. This procedure is perturbative
in the 2-particle interaction, but non-perturbative in the impurity
strength. 
Within our approximate scheme $\Sigma^{\Lambda}$ has no
imaginary part, which implies the vanishing of the vertex corrections 
to the conductance formula (\ref{conductf}).\cite{Oguri}
The Green function $\tilde G_{1,N}$ entering Eq.\ (\ref{conductf})
is obtained from the Dyson equation with $\Sigma^{\Lambda=0}$. 

As a test of our method at finite $T$ we first discuss results for 
the single barrier problem. We find the expected scaling behavior of the 
conductance (for an example see the solid line in the upper panel of  
Fig.\ \ref{fig1}).\cite{KaneFisher,VM1} From the suppression of 
$G(T) \propto T^{2 \alpha_B}$, the exponent $\alpha_B$ can be 
read off. Within our truncation scheme we obtain an approximation
$\alpha_B^{\rm RG}(U)$ for $\alpha_B(U)$. Consistent with the 
exponents of the $1/N$-scaling determined in 
Ref.\ \cite{VM1} at $T=0$ we find 
$\alpha_B^{\rm RG}(0.5)=0.165$, 
$\alpha_B^{\rm RG}(1)=0.35$, and $\alpha_B^{\rm RG}(1.5)=0.57$ 
in good agreement with the exact exponents 
$\alpha_B(0.5)=0.1608$, $\alpha_B(1)=1/3$, and  
$\alpha_B(1.5)=0.5398$ known from the Bethe
ansatz solution of the bulk model.\cite{Haldane}    

\begin{figure}[htb]
\begin{center}
\includegraphics[width=.4\textwidth,clip]{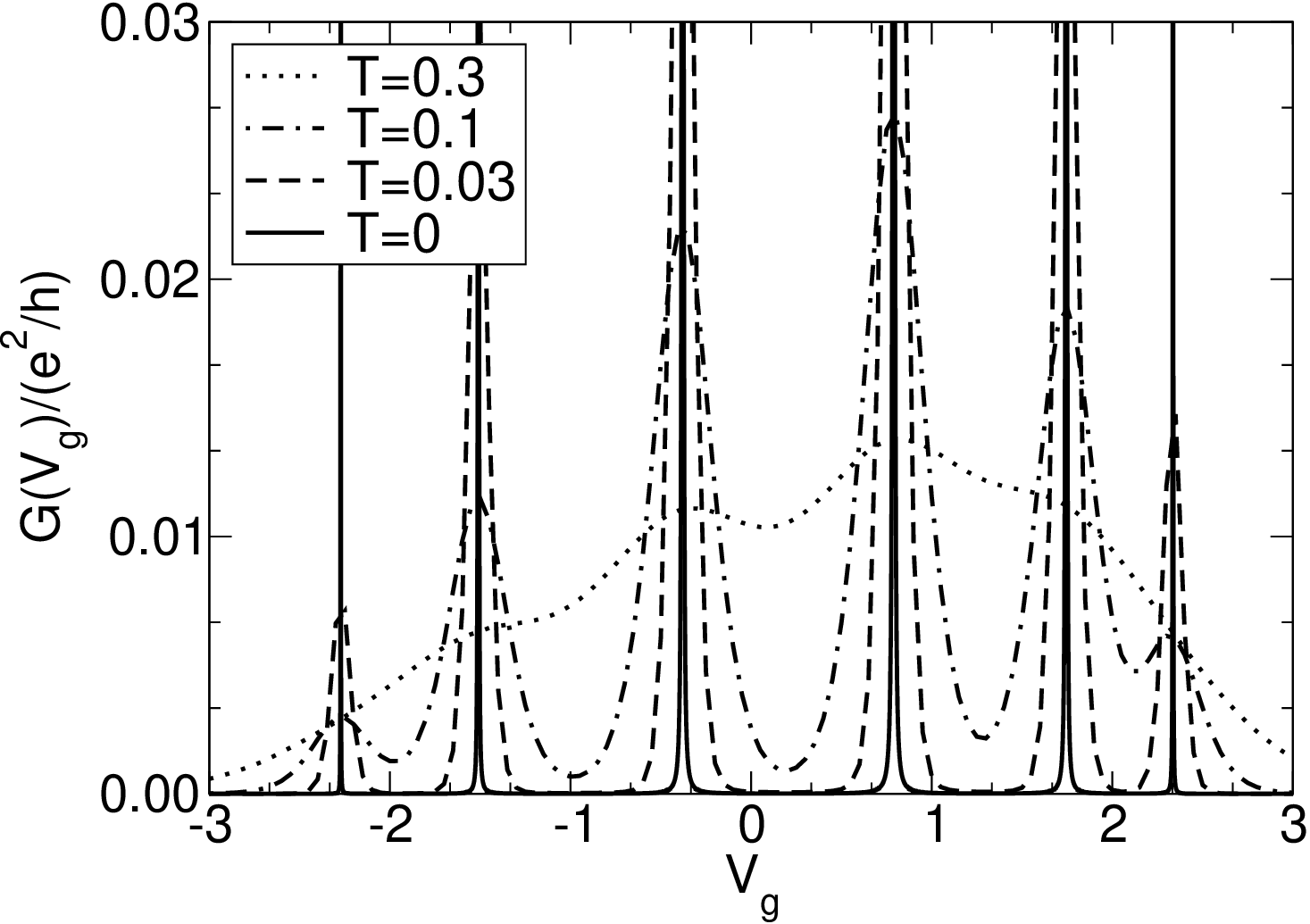}
\end{center}

\vspace{-0.7cm}

\caption[]{The conductance as a function of gate voltage for $N_D=6$,
  $U=0.5$, $V_{l/r}=10$, $N=10^4$, and different $T$.\label{fig0}}
\end{figure}

\begin{figure}[htb]
\begin{center}
\includegraphics[width=0.4\textwidth,clip]{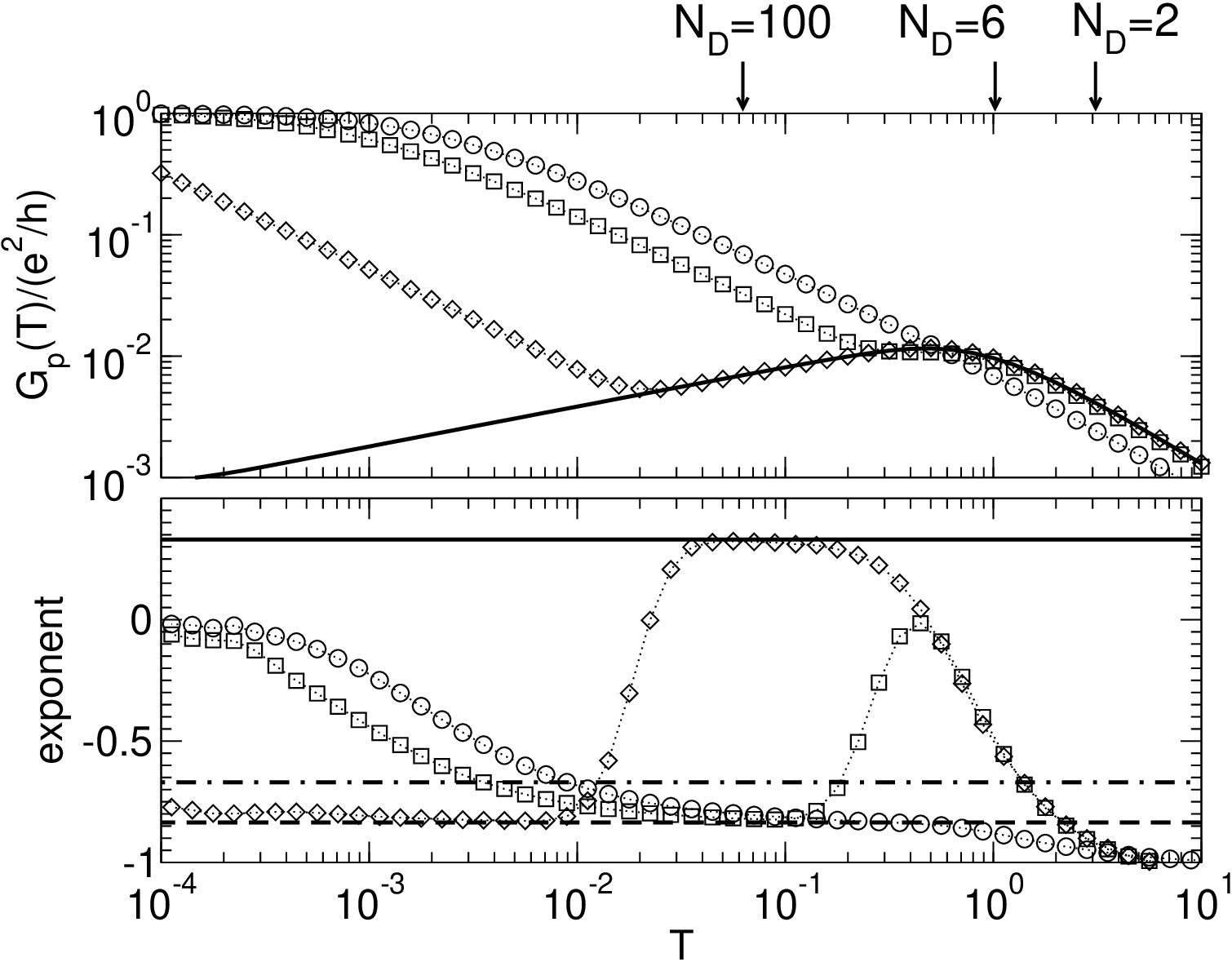}
\end{center}

\vspace{-0.7cm}

\caption[]{Upper panel: $G_p(T)$ for
  $U=0.5$, $N=10^4$, $V_{l/r}=10$, and 
  $N_D=2$ (circles), $6$ (squares),
  and $100$ (diamonds). The arrows indicate $\Delta_D$. 
  The solid curve shows $G(T)/2$ for a single
  barrier with $V=10$ and $U=0.5$, $N=10^4$. Lower panel: Logarithmic
  derivative of $G_p(T)$. Solid line: $2 \alpha^{\rm RG}_B$; dashed
  line: $\alpha_B^{\rm RG} -1$; dashed-dotted line: $2\alpha_B^{\rm RG}
  -1$.  \label{fig1}}
\end{figure}

For the double barrier problem we start the discussion with the case
of strong, symmetric barriers.  
In Fig.\ \ref{fig0} $G(V_g)$ is shown for $N_D=6$, $U=0.5$,
$V_{l/r}=10$ and different $T$. In all figures the 
wire length is chosen to be $N=10^4$ sites in 
rough agreement with the 
length of carbon nanotubes accessible to transport 
experiments.\cite{Postma} For sufficiently small $T$ 
resonance peaks are clearly developed. For $T \to 0$, $G_p(V_g^r)$
tends towards $e^2/h$. Because of $N < \infty$ even at 
$T=0$ the peaks have  finite width. For all peaks
the $T=0$ width scales to zero as 
$N^{K^{\rm RG}-1}$,\cite{KaneFisher} with the fRG approximation 
$K^{\rm RG}$ for $K$.   The different height
and width of the peaks at fixed $T$ is a finite band effect. 
In the following we always study the behavior 
of the resonance closest to $V_g=0$. We have verified that $G_p(T)$ 
for the other peaks shows a similar $T$-dependence.

In the upper panel of Fig.\ \ref{fig1} $G_p(T)$ is shown for  
$U=0.5$, $V_{l/r}=10$, and  $N_D=2$, $6$, $100$. 
The lower panel contains the logarithmic derivative of $G_p(T)$ 
from which the exponents of possible power-law behavior can be read 
off directly. Several temperature regimes can be distinguished. 
For $T$ larger than the bandwidth, $G_p(T) \propto T^{-1}$. 
For $T < \Delta_D$, marked by arrows in Fig.\ \ref{fig1}, down to a
lower bound $T^\ast$ discussed below we find 
$G_p(T) \propto T^{\alpha_B^{\rm RG}-1}$ 
as indicated in  the lower panel of Fig.\ \ref{fig1} by the dashed 
line. This is the regime of UST,\cite{Furusaki2}
in which the integral Eq.\ (\ref{conductf}) leading to $G_p(T)$ 
is dominated by a single peak of $|t(\varepsilon,T)|^2$, but $T$ is still
much larger than the width of this peak. 
Peaks in $G_p(V_g)$ are clearly separated and the width $w(T) \propto
T$ can be read off. The size of the UST regime grows with increasing
$N_D$ and shifts towards smaller $T$. 
For $N_D \geq 4$ we find non-monotonic behavior of $G_p(T)$.
For $N_D \gtrapprox 30$ the region of decreasing conductance 
(with decreasing $T$) can be described by a power-law   
with exponent $2 \alpha_B^{\rm RG}$ (solid line in the lower panel of
Fig.\ \ref{fig1}).\cite{Furusaki1} For these $T$ many peaks of 
$|t(\varepsilon,T)|^2$ contribute to the integral.
For $V_{l/r} \gg 1$ and $N_D \gg 1$, i.e.~in the parameter regime with
distinctive power-law scaling, the non-monotonic behavior can be 
understood using Kirchhoff's law. The solid 
line in the upper panel of Fig.\ \ref{fig1} displays $G(T)/2$ obtained 
for a single barrier with $V=10$ and $U=0.5$. The 
comparison shows that the conductance of the double barrier
problem for $T$ up to the local minimum can be 
determined by adding up resistances of the related single barrier
case. We do not observe indications of CST with exponent $2
\alpha_B^{\rm RG}-1$, shown as the dashed-dotted line in the lower
panel of Fig.\ \ref{fig1}. 
For $T \to 0$, $G_p(T)/(e^2/h)$ saturates at $1$. 
Due the finiteness of the wire for $T  \ll \Delta_W$, with $\Delta_W =
\pi v_F/N$, 
we find $1-G_p(T)/(e^2/h) \propto T^2$. For small $N_D$ and
intermediate to small barriers  
at increasing $T$ this behavior is followed by a regime with 
$1-G_p(T)/(e^2/h)\propto T^{2 K^{\rm RG}}$.\cite{Furusaki1}
All the above scaling regimes we also find for $U=1$ and $1.5$.

Temperature regimes with power-law behavior
and the same scaling exponents were obtained using a field theoretical
model and perturbation theory in $V_{l/r}$ as well as
$1/V_{l/r}$.\cite{Furusaki1,Furusaki2} The agreement of the results
obtained from two complementary approximate methods and for different
models constitutes a strong argument for the correctness of the
physics obtained.  
These considerations put very tight limits on  
the parameters for which a scaling regime with exponent 
$2 \alpha_B-1$ could occur.\cite{Thorwart}
As an advantage of our method results for $G_p$ on all energy 
scales down to $T=0$ can be obtained within one approximation scheme
and for more realistic lattice models. Furthermore the fRG allows for
extensions to non-perfect contacts which have to be considered 
when comparing with experiments.\cite{Postma}

\begin{figure}[htb]
\begin{center}
\includegraphics[width=0.4\textwidth,clip]{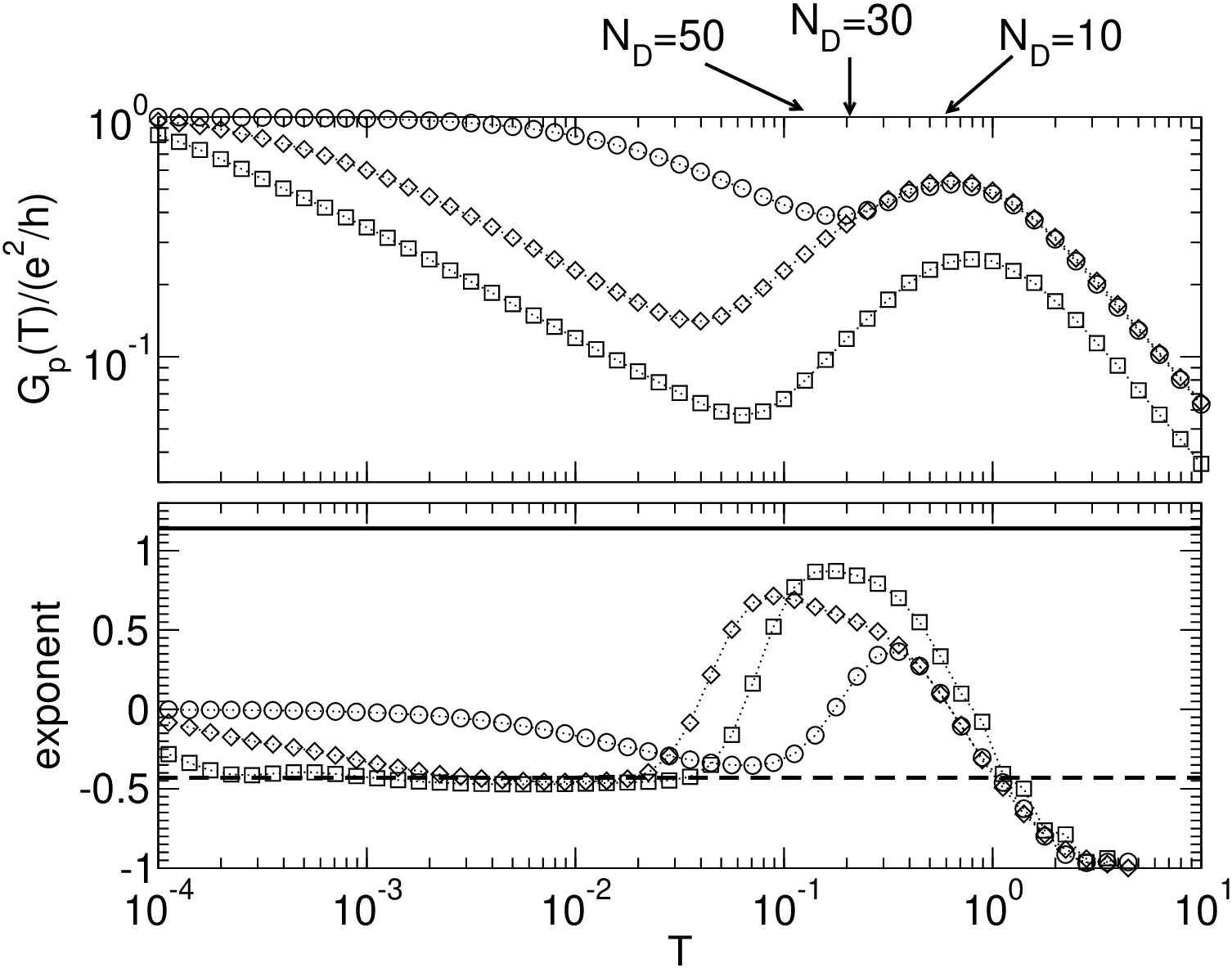}
\end{center}

\vspace{-0.7cm}

\caption[]{Upper panel: $G_p(T)$ for
  $U=1.5$, $N=10^4$,  
  $N_D=10$  with $V_{l/r}=0.8$ (circles), $N_D=30$  with $V_{l/r}=1.5$
  (squares), and $N_D=50$  with $V_{l/r}=0.8$ (diamonds). The arrows 
  indicate $\Delta_D$.
  Lower panel: Logarithmic
  derivative of $G_p(T)$. Solid line: $2 \alpha^{\rm RG}_B$; dashed
  line: $\alpha_B^{\rm RG} -1$. \label{fig4}}
\end{figure}

We next discuss the case of weak to intermediate barriers, dots of
a few ten lattice sites, and larger $U$. In this part of the parameter
space we find non-universal behavior of $G_p(T)$ which so far has not
been studied theoretically and provides an alternative interpretation
of the QMC data.\cite{Huegle} It might also be relevant in connection 
with experiments\cite{Postma} since it is not clear that 
the experimental dot parameters fall in the regime of 
universal scaling.  
In Fig.\ \ref{fig4} $G_p(T)$ is shown for $U=1.5$, $N_D=10$ with
$V_{l/r}=0.8$ (circles), $N_D=30$ with $V_{l/r}=1.5$ (squares), and
$N_D=50$ with $V_{l/r}=0.8$ (diamonds). These parameter sets are 
chosen to be close to the set used in the QMC 
calculations.\cite{Huegle} 
In the limit of weak barriers and small dots (circles) neither the power-law
with the exponent $2 \alpha_B$
nor the one with $\alpha_B-1$ are clearly developed. 
For increasing $N_D$ or increasing $V_{l/r}$ the UST regime 
gets more pronounced. For weak to intermediate barriers no plateau 
with exponent $2 \alpha_B$ is established (see the lower panel). 
The behavior in the regime of
decreasing $G_p(T)$ (for decreasing $T$) depends on $V_{l/r}$ and $N_D$
and is thus non-universal. It might be identified incorrectly 
as a power-law with 
an exponent significantly smaller than $2 \alpha_B$, i.e. in the
vicinity of  $2 \alpha_B-1$. 

Our fRG results for $G_p(T)$ in the case of asymmetric
barriers and the off resonance scaling of $G(T)$ for both symmetric
and asymmetric barriers will be presented in Ref.\ \cite{Tilman}.
For later reference we mention that in all these cases  
we obtain $T^{2 \alpha_B^{\rm RG}}$ for $T \to 0$  as 
expected.\cite{KaneFisher}

Using the projection operator onto the subspace of the one-particle
states of the dot and the orthogonal complement to the left and right
of it the result 
 $|t(\varepsilon,T)|^2 = (4- \varepsilon^2) 
|\tilde G_{1,N}(\varepsilon,T)|^2$ can be written in a 
form which is useful for obtaining a deeper 
understanding of the different scaling regimes discussed above and to
work out the differences of our approach to the ``leading-log''
approximation.
As an example we take the case of hopping barriers
and $N_D=1$. Then one obtains
\begin{eqnarray}
&& |t(\varepsilon,T)|^2 = \label{formula}\\ && \frac{4 \Gamma_l(\varepsilon,T)
  \Gamma_r(\varepsilon,T)}{\left[ \varepsilon - V_g - 
  \Omega_l(\varepsilon,T) - \Omega_r(\varepsilon,T)\right]^2 + \left[ 
  \Gamma_l(\varepsilon,T) + \Gamma_r(\varepsilon,T) \right]^2} 
\nonumber
\end{eqnarray}
with $\Gamma_{l} = t_{l}^2 \, \mbox{Im} \, \tilde
G_{j_l-1,j_l-1}^0$, $\Omega_l =
t_l^2 \, \mbox{Re} \,  \tilde G_{j_l-1,j_l-1}^0$
and similar expressions for $l \to r$. The Green function $\tilde
G_{i,j}^0(\varepsilon,T)$
is obtained by considering $\Sigma^{\Lambda=0}$ as an effective
potential outside the dot and setting $t_{l/r}=0$. 

For $V_g=0$, $\varepsilon - V_g - \Omega_l(\varepsilon,T) - 
\Omega_r(\varepsilon,T)$ vanishes at $\varepsilon=0$ and we find a 
resonance. Due to the generalized Breit-Wigner form of 
Eq.\ (\ref{formula}) for $t_l=t_r$ the transmission at 
resonance is 1 independent of the dependence of
$\Gamma_{l/r}$ on its arguments and the parameters. It is important to 
note that in this case and for $T \to 0$,
(i) $\Gamma_{l/r}(\varepsilon=0,T)$ does not follow the 
single barrier scaling $T^{\alpha_B}$ but instead saturates 
(even for $N \to \infty$). For asymmetric barriers 
with $t_r <t_l <1$ and at resonance $\Gamma_{r}(\varepsilon,T) 
\propto T^{\alpha_B^{\rm RG}}$ while 
(ii) $\Gamma_{l}(\varepsilon,T)$ increases as 
$T^{-\alpha_B^{\rm RG}}$. Similar scaling is found for the
$\varepsilon$ and $1/N$ dependence. These power-laws are cut-off 
by the largest of the scales $T$, $\varepsilon$, and $\Delta_W$. 
Inserting the expressions for $\Gamma_{l/r}$ into
Eq.\ (\ref{formula}) leads to a power-law suppression of 
$|t(\varepsilon,T)|^2$ with exponent $2 \alpha_B^{\rm RG}$ which 
directly gives $G_p(T) \propto T^{2 \alpha_B^{\rm RG}}$ mentioned 
above. We next study the off resonance case with $V_g \neq 0$ and 
$t_l=t_r \leq 1$ including the single barrier situation ($t_{l/r}=1$). 
Because of the symmetry in the following we suppress the indices $l/r$.
We focus on $T=0$ and consider $\Delta_W \propto 1/N$ as the relevant energy
scale. Then the integral in Eq.\ (\ref{conductf}) can be performed
and $|t(0,0)|^2$ is directly related to $G$. For 
small $V_g$ we (iii) find $\Gamma(0,0) \propto \mbox{const.}$ and $V_g + 
2 \Omega (0,0) \propto V_g N^{1-K^{\rm RG}}$, which combines to 
the expected $1- G/(e^2/h) \propto N^{2(1-K^{\rm RG})}$. 
For large $V_g$, (iv) $\Gamma(0,0) \propto t^2 N^{-\alpha_B^{\rm RG}}$ and 
$V_g + 2 \Omega (0,0) \propto V_g$ which leads to  $G \propto N^{-2
  \alpha_B^{\rm RG}}$. The observations (i)-(iv) are not captured by
the parameterization of the effective transmission
$|t(\varepsilon,T)|^2$ used in the ``leading-log'' 
method.\cite{Nazarov,Polyakov,Glazman} In this approach the important 
energy dependence of the real part $\Omega$ [see (iii) and (iv)] 
does not appear.

Using Eq.\ (\ref{formula}) and its generalization for $N_D >1$\cite{Tilman}  
the UST regime can be understood quite simply.\cite{footnote-1}  
We here focus on symmetric barriers. 
Approaching from large $T$ UST sets in at a scale at which only one peak 
of $|t(\varepsilon,T)|^2$ contributes to the integral in Eq.\ 
(\ref{conductf}) 
(respectively for $T$ smaller than the bandwidth at $N_D=1$). 
As can be seen in Fig.\ \ref{fig1} this scale is significantly smaller
than $\Delta_D$. For these $T \gg | \varepsilon |$ we find 
$\Gamma (\varepsilon,T) \propto t^2 T^{\alpha_B^{\rm RG}}/N_D$. 
Performing the integral in Eq.\ (\ref{conductf}) leads to $G_p(T) \propto
T^{\alpha_B^{\rm RG}-1}$ and $w(T) \propto T$. 
This scaling continues down to the temperature $T^\ast \propto
(t^2/N_D)^{1/(1-\alpha_B^{\rm RG})}$ at which the width 
of the peak in $|t(\varepsilon,T)|^2$
is equal to $T$. This defines the lower bound of the UST regime.   

In conclusion, applying the fRG 
we obtained a comprehensive picture
of resonant tunneling in a microscopic model for a Luttinger liquid
with $1/2 \leq K \leq 1$.
Depending on the dot parameters we found distinctive power-law 
scaling of $G_p(T)$, but also regimes with non-universal behavior. No 
signatures of CST were found.    

We thank X.~Barnab\'e-Th\'eriault, R.~Egger, M.~Grifoni, A.~Komnik,
D.~Polyakov, M.~Thorwart, and especially H.~Schoeller for
discussions.


\begin{thebibliography}{*}
\bibitem{footnote-1} In contrast to what 
is found in the model of Ref.\ \cite{Komnik} for not too small
barriers we observe indications of UST with exponent $\alpha_B-1$ 
also for $N_D=1$. At $K=1/2$, i.e.\ $\alpha_B=1$ this leads to a 
plateau in $G_p(T)$.
\end{thebibliography}
\end{document}